\documentclass[prl,twocolumn,english,floatfix]{revtex4}

\usepackage{babel}
\usepackage{graphicx}
\usepackage{bm, amsmath, amssymb}
\usepackage{xcolor}
\usepackage{color}
\usepackage{soul}
\usepackage{pdfpages}
 \usepackage{babel}
\usepackage{amstext}
\usepackage{bbold}
\usepackage{esint}
\usepackage[pdftex, pdftitle={Article}, pdfauthor={Author}]{hyperref} 

\begin{document}

\title{Broadband Microwave Isolation with Adiabatic Mode Conversion in Coupled Superconducting Transmission Lines}






\author{Mahdi Naghiloo$^{1}$}
\author{Kaidong Peng$^{1,2}$}
\author{Yufeng Ye$^{1,2}$}
\author{Gregory Cunningham$^{1,3}$}
\author{Kevin P. O'Brien$^{1,2}$}

\email[Correspondence email address: ]{kpobrien@mit.edu}%

\affiliation{$^1$Research Laboratory of Electronics, Massachusetts Institute of Technology, Cambridge, MA 02139, USA}
\affiliation{$^2$Department of Electrical Engineering and Computer Science, Massachusetts Institute of Technology, Cambridge, MA 02139, USA}
\affiliation{$^3$Harvard John A. Paulson School of Engineering and Applied Sciences, Harvard University, Cambridge, MA 02138, USA}


\begin{abstract}
We propose a traveling wave scheme for broadband microwave isolation using parametric mode conversion in conjunction with adiabatic phase matching technique in a pair of coupled nonlinear transmission lines. This scheme is compatible with the circuit quantum electrodynamics architecture (cQED) and provides isolation without introducing additional quantum noise. We first present the scheme in a general setting then propose an implementation with Josephson junction transmission lines. Numerical simulation shows more than 20 dB isolation over an octave bandwidth (4-8\,GHz) in a 2000 unit cell device with less than 0.05 dB insertion loss dominated by dielectric loss.
\end{abstract}

\maketitle



\emph{Introduction}---Non-reciprocal devices are essential for protecting quantum systems from their noisy electromagnetic environments while allowing measurement and control. Superconducting circuits, one of the leading platforms for realizing quantum computers \cite{arute2019quantum}, currently uses commercial bulky isolators which utilize permanent magnets to realize uni-directional propagation of microwave signals \cite{pozar2009microwave,Fay1965OperationOT}. However, the goal of developing useful quantum computers with potentially millions of qubits \cite{national2019quantum,preskill1998reliable} demands integration of the essential components and dramatic reductions in the size of the supporting hardware. These challenges in superconducting circuits as well as other quantum systems, motivate the search for miniaturized on-chip alternatives for ferromagnetic isolators utilizing various physics such as acousto-optics \cite{kang2011reconfigurable}, electromechanics \cite{barzanjeh2017mechanical}, non-Hermitian dynamics \cite{ramezani2010unidirectional,bender2013observation} and nonlinear parametric processes \cite{kamal2011noiseless, chapman2019design,chapman2017widely, ranzani2017wideband,sounas2017non}.
In particular, parametric processes in traveling wave devices are encouraging for implementing non-reciprocal dynamics as the required phase-matching conditions are usually met only in one propagation direction. Especially, parametric frequency and mode conversions are promising for implementing isolation for high efficiency and high fidelity measurements because these processes in principle can be noiseless in contrast with parametric amplification or non-Hermitian dynamics where the gain and loss introduce inevitable noise into the system \cite{haus1962quantum,caves1982quantum,clerk2010introduction}. Several theoretical and experimental works have studied the isolation using parametric frequency and mode conversions \cite{yu2009complete,lira2012electrically,ranzani2017wideband,doerr2014silicon}; however, the experimental realization of a parametric device that provides sufficient isolation over a broad instantaneous bandwidth has remained elusive. Here we present a practical scheme for broadband isolation based on adiabatic phase-matched parametric mode conversion. We propose a realization using coupled transmission lines in superconducting quantum circuits, but it is important to note that our scheme is general and applicable to a variety of platforms. Our circuit-level analysis with realistic component values suggests more than 20 dB isolation over an octave of bandwidth, comparable to commercial isolators, but, with a smaller footprint and a superconducting qubit compatible fabrication process.

\begin{figure*}[t!]
\centering
\includegraphics[width=0.93 \linewidth]{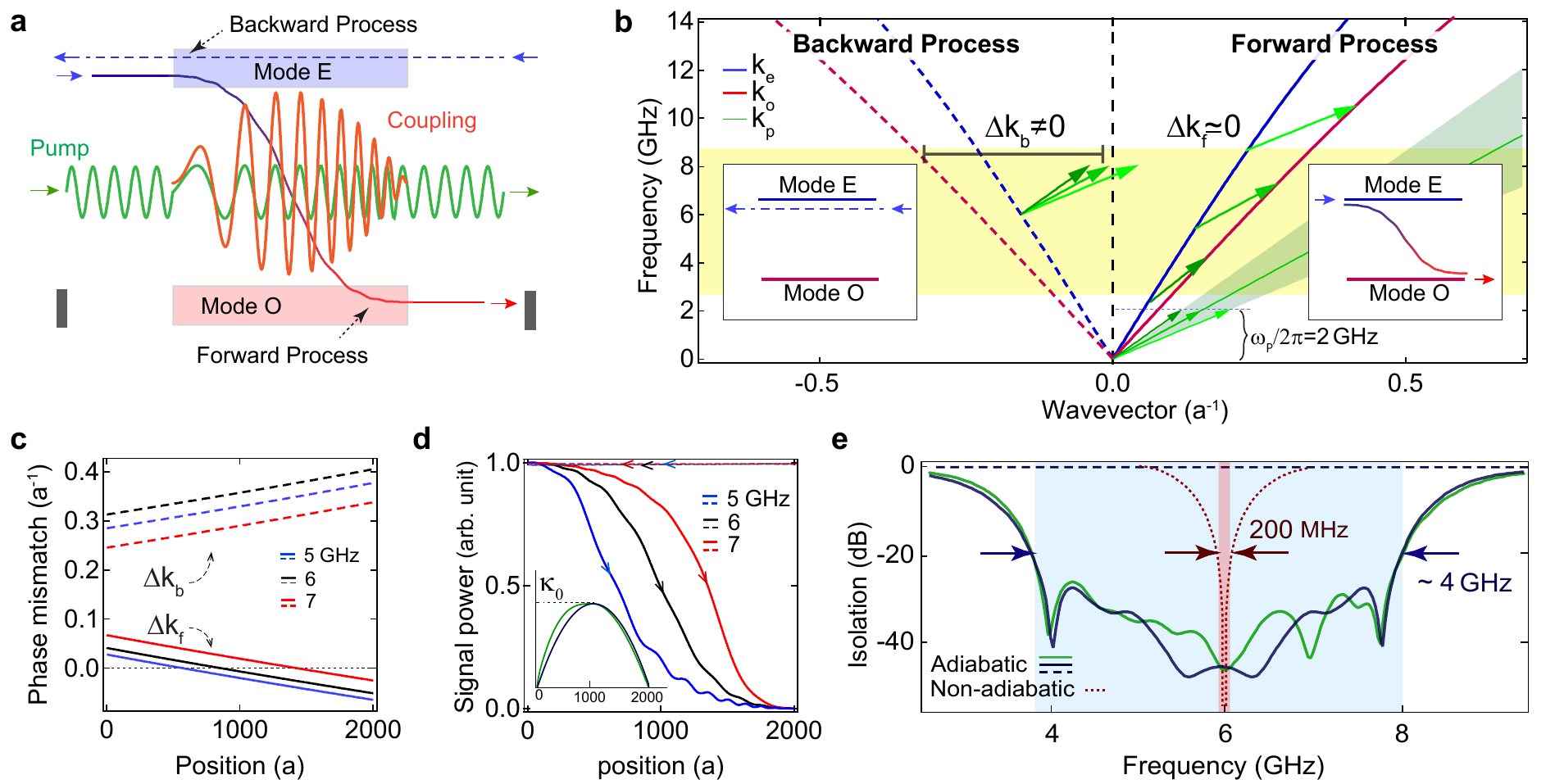}
\caption{ \textbf{Isolation through adiabatic mode conversion}: \textbf{a.} Modes E and O represent even and odd propagating modes in a waveguide. The coupling between two modes (shown in orange) is spatially varied both in terms of wavevector and strength via a pump mode shown in green. The conversion of mode E to mode O happens only in the forward direction. \textbf{b.} The blue (red) solid line is the dispersion relation for mode E (mode O). The dashed lines correspond to backward propagation. The green solid line shows pump dispersion relation which varies within the green-shaded region. The green arrow shows wavevectors correspond to 2 GHz pump tone which connects mode E and mode O within a highlighted range of frequency only in the forward direction but poorly phase-matched in the backward direction. \textbf{c.} Forward/backward phase mismatch for three different signal frequencies. \textbf{d.} Solid (dashed) lines show the adiabatic conversion of signal in the forward (backward) direction for three different input signal frequencies. The inset shows the coupling spatial variation (see Eq.\,\ref{quadratic_ramp}) \textbf{e.} The Blue solid line shows the isolation based on our presented scheme.  The green line is the isolation performance of the scheme implemented in cQED which provides more than 20 dB isolation over 4 GHz range of frequency (shaded region in blue). The dashed line on top is the transmission of the backward signal. The red dotted line is the isolation performance without adiabatic conversion which provides only 200 MHz bandwidth of isolation (shaded region in red). \label{fig1} }
\end{figure*}



\emph{Adiabatic parametric mode conversion}---Parametric mode conversion provides unidirectional conversion of the signal between two modes of propagation. This nonreciprocal dynamics for the signal occurs because, with a preferred propagation direction set by pump tone, the phase-matching condition for the parametric process is met only in one direction \cite{o2014resonant}. This property naturally suggests utilizing parametric conversions to implement isolator, which has been studied in optics and microwave in traveling wave devices \cite{yu2009complete,lira2012electrically,ranzani2017wideband}. 
However, the major limitation of the isolation based on parametric conversions is that the full conversion only occurs around a certain frequency set by the device characteristics (e.g. device length, coupling strength, pump power). Therefore the isolation based on parametric mode conversion would not be broadband \cite{yu2009complete,lira2012electrically} and the operating frequency is sensitive to parameter variations. However, as we propose in this letter, by combining adiabatic techniques \cite{suchowski2008geometrical,suchowski2009robust,suchowski2014adiabatic} with parametric mode conversion,  we can realize broadband conversion and thus broadband isolation.

In Fig.\,\ref{fig1}a, we schematically illustrated the proposed adiabatic mode conversion. Here, mode E and O represent two transmission lines or two orthogonal modes of propagation in a waveguide with different dispersion relations as illustrated in Fig.\,\ref{fig1}b. The propagating signals in forward/backward direction in mode E and O can be represented as $E_{f/b}(x,t)=E_{f/b}(x) e^{-i(\omega_e t \mp k_e x)}+ \mathrm{c.c.}$ and $O_{f/b}(x,t)=O_{f/b}(x) e^{-i(\omega_o t \mp k_o x)}+\mathrm{c.c.}$, where $|E_{f/b}(x)|^2$ and $|O_{f/b}(x)|^2$ quantify the average number of photons propagating in the forward/backward direction in each mode. In the normal situation, mode E and O are decoupled and the incoming signal in each mode propagates through the medium without conversion. However, we consider a situation in which a propagating coherent pump tone in the third mode induces coupling between mode E and O in the form of 
\begin{equation}\label{pump_couplin_form}
\kappa(x,t)=\frac{1}{2}\kappa(x) \exp[ \ -i\omega_p t +i \int_0^x k_p(x) dx \ ]+ \mathrm{c.c.}
\end{equation}
Here $\omega_p=\omega_o- \omega_e$ is the pump frequency which is fixed unlike the pump wavevector
\begin{equation}\label{LCS}
 k_p(x)=K + \frac{2 \alpha}{L} (x-\frac{L}{2}),
\end{equation}
and the coupling strength
\begin{equation}\label{quadratic_ramp}
\kappa(x)= \frac{4 \kappa_0}{L^2} (L-x)x,
\end{equation}
which are slowly-varying functions of position. Here we choose a linear variation for pump wavevector throughout the device which is quantified by pump wavevector value at the center of the device $K$ and its variation range $\alpha$. The induced coupling strength has quadratic dependence on position and attains its maximum $\kappa_0$ at the center and vanishes at both ends of the device.

The dynamics in the forward direction of propagation for signals in mode E and O can be described by the following coupled mode equations,
\begin{eqnarray}\label{ode00}
\frac{d}{dx} \begin{pmatrix}
E_f\\
O_f 
\end{pmatrix}= \frac{i}{2}\begin{pmatrix}
- \Delta k_f(x) &   \kappa(x) \\
 \kappa(x) &  \Delta k_f(x) 
\end{pmatrix} . \begin{pmatrix}
E_f\\
O_f 
\end{pmatrix},
\end{eqnarray}
where $\Delta k_f(x)=k_o - k_e - k_p(x)$ is the instantaneous (position dependent) phase mismatch in the forward direction.
In our simple model, the forward and backward dynamics are decoupled and similar equations hold for the backward process, except the mismatch is given by  $\Delta k_b(x) = k_o - k_e + k_p(x)$. 
We note that the dynamical matrix in Eq.\,\ref{ode00} is analogous to a driven qubit Hamiltonian ($x\rightarrow t$) with variable detuning $\Delta k_f(x)$ and Rabi drive $\kappa(x)$ \cite{oliver2005mach}.

Give the variation of the pump wavevector $k_p(x)$, the pump tone connects (perfectly phase matches) different frequencies in mode E to mode O at different positions for forward propagating signals.
For example, as illustrated in Fig.\,\ref{fig1}b, here we consider a 2 GHz pump tone (green arrows) that couple mode E to mode O from 2.7 GHz at $x=0$ (dark green arrow) to 9.3 GHz at $x=L$ (light green arrow).
Assuming linear dispersion relations, this results in a spatially linear variation for mismatch $\Delta k_{f/b}(x)$ (Fig.~\ref{fig1}c). With a proper choice of the variation parameter $K$ and $\alpha$, the mismatch $\Delta k_{f}(x)$ starts from a non-zero value and crosses zero during signal propagation in the forward direction as depicted in Fig.~\ref{fig1}c. Therefore, the dynamics in the forward direction would be ``spatial" version of rapid adiabatic passage in spin physics or Landau-Zener technique in quantum two-level systems \cite{zener1932non,oliver2005mach,shevchenko2010landau}.
At the same time, the mismatch in backward direction $\Delta k_{b}(x)$ stays far from zero throughout the signal propagation in backward direction (Fig.~\ref{fig1}c). With the same analogy to two-level system, the dynamics in backward direction would be ``spatial" version of a far-detuned driven qubit (e.g. $[ \Delta k_b(x)/\kappa(x)]^2 \gg 1$).
Therefore, we have adiabatic conversions between two modes only in the forward direction of propagation and negligible conversions in backward direction.
Note that, the effect of the coupling strength variation is to improve the adiabatic conversion efficiency in close connection to pulse shaping techniques in NMR \cite{melinger1994adiabatic, garwood2001return, goswami2003optical}, cold atoms \cite{du2016experimental}, and qubit gate optimization \cite{martinis2014fast,guerin2011optimal,gambetta2011analytic,yang2017achieving}. Except, in our case, the variation occurs in position and the objective is to have a robust conversion for a broad range of frequencies.
Both spatial variations of the pump wavevector and amplitude (strength) associated with the coupling are schematically illustrated in Fig.\,\ref{fig1}a.

The adiabatic dynamics governed by Eq.\,\ref{ode00} can be solved numerically as depicted in Fig.\,\ref{fig1}d for three different incoming signals in forward and backward direction in mode E using realistic parameters (per unit length $a$) in cQED $\kappa_0=0.023 \, a^{-1}$, $\alpha=0.05\, a^{-1}$ and $L=2000 a$. 
 As depicted in Fig.~\ref{fig1}d, the signal starting in mode E at $x=0$ is efficiently converted into mode O through an adiabatic conversion while the conversion in the backward direction is negligible.
 More importantly, this nonreciprocal dynamics happens over a wide bandwidth of frequency for the input signal. With terminations on both sides of mode O, this configuration acts as a two-port broadband isolator. The negligible conversion in the backward propagation contributes to the total insertion loss which is calculated to be less that 0.05\,dB with dielectric loss tangent $\delta=10^{-5}$ \cite{o2008microwave}. The broadband isolation performance of this scheme is depicted in Fig.\,\ref{fig1}e where the blue solid (dashed) line shows isolation (transmission) for the given parameters.  For comparison, the red dotted line shows the corresponding isolation without performing spatial variation of the pump wavevector and coupling strength. The green line is the isolation performance of a practical implementation of the presented scheme in cQEC which is the topic of our discussion in the rest of this letter.

\begin{figure}[]
\centering
\includegraphics[width=0.98 \linewidth]{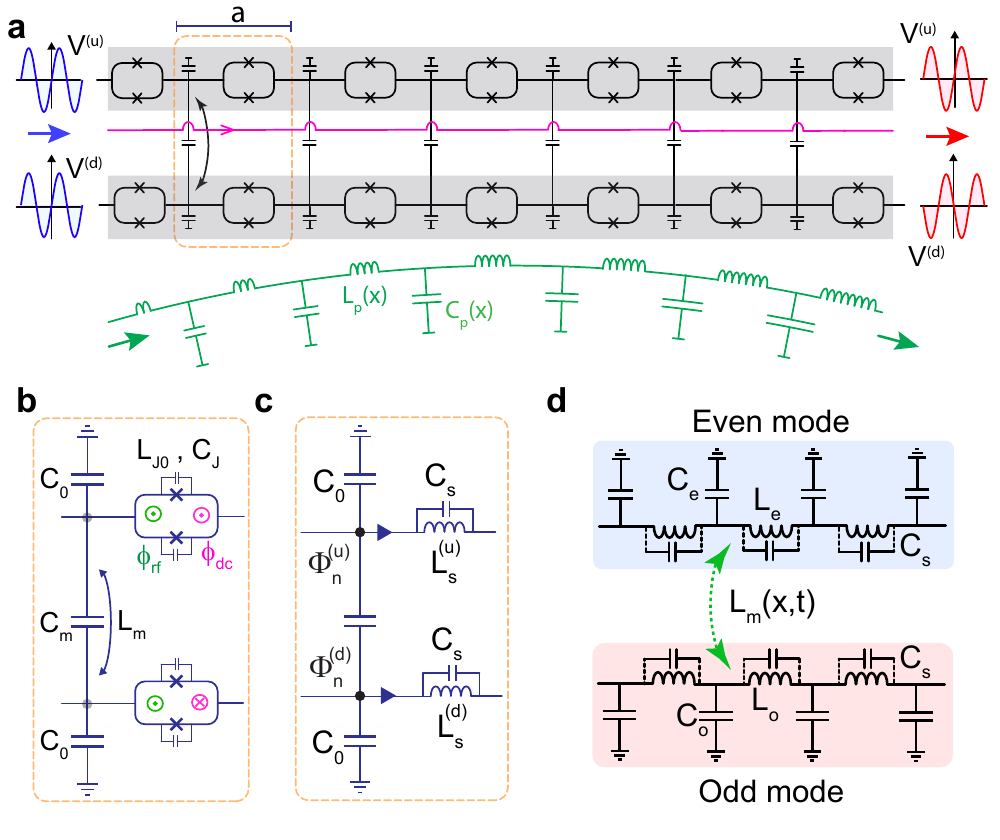}
\caption{\textbf{Circuit diagram:} \textbf{a.} The device consists of a pair of coupled identical transmission lines (shaded). The pump transmission line (in green) provides in-phase flux modulation $\phi_{rf}$ for SQUIDs in both transmission lines. The rf flux amplitude is spatially controlled as schematically illustrated. The dc flux (provided by the red line) has opposite direction in the upper and lower transmission line. \textbf{b.} The flux configuration in a unit cell. Here, $C_m$ and $L_m$ are geometric capacitance and inductive coupling per unit cell between two transmission lines. \textbf{c.} SQUIDs can be considered as inductors with flux-modulated inductance. The inductance is modulated in opposite directions (Eq.\,\ref{eq_ind}).  \textbf{d.} The equivalent circuit of a coupled transmission line in even and odd mode basis. Even and odd modes are normally orthogonal, except, the modulated flux induces an effective inductive coupling between two modes by breaking the symmetry between upper and lower transmission lines.\label{fig2}} 
\end{figure}
\emph{Implementation in superconducting circuit}---Now, we proceed by discussing a practical implementation of this scheme in the superconducting circuit platform. The implementation has four essential components: I. Realizing two orthogonal modes of propagation (mode E and mode O) with \textit{distinct} dispersion relations; II. Implementing the coupling between these two modes in a traveling wave fashion (Eq.\,\ref{pump_couplin_form}); III. Carrying out the pump wave vector sweep (adiabatic phase matching) (Eq.\,\ref{LCS}); IV. And finally implementing the coupling strength variation (Eq.\,\ref{quadratic_ramp}).

As depicted in Fig.\,\ref{fig2}a, we consider a pair of coupled lumped element transmission lines where the inductance in each unit-cell is provided predominantly by a pair of Josephson junctions in SQUID geometry. This forms a waveguide that supports two orthogonal modes of propagation; namely \textit{even} and \textit{odd} modes. The effective inductance of a SQUID is tuned by a dc bias line (shown in magenta) and also modulated by an rf pump in a separate transmission line (shown in green).
We consider the configuration in which SQUIDs in the upper and lower lines receive dc flux in opposite directions while the rf flux modulations are in-phase for both lines as depicted in Fig.\,\ref{fig2}b. Therefore the effective inductance of SQUIDs in upper and lower transmission line would be,
\begin{eqnarray}
L_s^{\mathrm{u,d}}=0.5 L_{J0} / \mid \cos \left(\frac{ \pm \phi_{dc} + \phi_{rf}(x,t)}{2 \phi_0}\right) \mid \, ,
\end{eqnarray}
where $L_{J0}$ is the bare inductance of a single Josephson junction and  $\phi_0=\Phi_0/2\pi$ is the reduced flux quantum and $\phi_{dc}$ is the magnitude of the dc flux threading each SQUID. The rf flux $\phi_{rf}(x,t)$ has both temporal and spatial dependence inherited from the pump transmission line. We consider the regime where the signal currents propagating in the coupled transmission lines are much smaller than the critical current of the Josephson junctions ($I \ll I_c$). In this regime, nonlinear effects due to signal propagation in the transmission lines are negligible. Therefore, SQUIDs can be considered as linear inductors whose inductance is modulated by $\phi_{rf}(x,t)$ (Fig.\,\ref{fig2}c). In the limit of small modulation, $\phi_{rf}(x,t)\ll \phi_{0}$ the effective inductance of SQUIDs in the upper and lower transmission line can be expressed as,
\begin{eqnarray}\label{eq_ind}
L_s^{\mathrm{u,d}}= \frac{L_{dc}}{1 \pm m(x,t)},
\end{eqnarray}
where the unitless parameter $m(x,t)=\tan(\frac{\Phi_{dc}}{2 \phi_0}) \frac{\phi_{rf}(x,t)}{2 \phi_0}$ accounts for flux-pump-induced modulation and $L_{dc}=0.5 L_{J0}/\mid \cos(\frac{\phi_{dc}}{2\phi_0}) \mid $ is the dc-biased SQUID inductance \cite{zorin2019flux}.

The nonlinear dynamics of the system can be described by Lagrangian approach by defining node-fluxes in each transmission line \cite{yaakobi2013parametric,zorin2019flux} as depicted in Fig.\,\ref{fig2}c. However, we are interested in the dynamics of the even and odd modes corresponding respectively to symmetric $\Phi^{(e)}_n = (\Phi^{(\mathrm{u})}_n +\Phi^{(\mathrm{d})}_n )/\sqrt{2}$ and anti-symmetric $\Phi_n^{(o)} = (\Phi^{(\mathrm{u})}_n -\Phi^{(\mathrm{d})}_n )/\sqrt{2}$ node-flux superposition in the two transmission lines \cite{orfanidis2002electromagnetic}. The equivalent circuit diagram in the even-odd basis is illustrated in Fig.\,\ref{fig2}d. Here $L_e=L_{dc}+L_m$, $L_o=L_{dc}-L_m$, $C_e=C_0$ and $C_o=C_0+2C_m$ \cite{orfanidis2002electromagnetic,supp}. In this basis, we have two otherwise orthogonal propagation modes where an effective inductive mode coupling $\mathbb{L}_m(x,t)= L_{dc}\, m(x,t)$ is induced by breaking the symmetry of upper and lower transmission lines via the rf flux.
Far below the cut-off frequency of the transmission lines $\omega_{0i}=1/\sqrt{C_i L_i}$, and the Josephson plasma frequency $\omega_{Ji}=1/\sqrt{C_s L_i}\,,\, i\in\{e,o\}$, the even and odd modes' characteristic impedance has the following form $Z_i=\sqrt{L_i/C_i}$ and the corresponding dispersion relations are,
\begin{eqnarray}\label{keko}
k_i(\omega)&=& \frac{\omega/\omega_{0i}}{\sqrt{1- \omega^2/\omega^2_{Ji}}}\simeq \sqrt{L_i C_i} \, \omega \, , \  i\in\{o,e\}.
\end{eqnarray}
In Fig.\,\ref{fig1}b, we plotted the dispersion curves for even/odd modes (blue/red lines) given a set of circuit parameters listed in Table~1. Note that the distinct dispersion relation for mode E and O is crucial for inhibiting the cascading parametric processes which otherwise would degrade the device performance\cite{yu2009complete}.
\begin{center}
\begin{table}[]
\resizebox{\linewidth}{!}{
\begin{tabular}{c|c|c|c}
\multicolumn{4}{c}
{Table~1: Circuit Parameters}\\
\hline
$C_0=80$ fF & $C_s=40$ fF& $C_m=20$ fF& $\phi_{dc} = 2 \pi/3 \ \phi_0$ \\
$L_{J0}=250$ pH & $L_m={\text -} \, 50$ pH & $m_0=0.1$ &$\phi_{rf}^{max} = 0.1\  \phi_0$\\
\end{tabular}}
\end{table}
\end{center}
In the continuum limit where unit cell length $a$ is much smaller than the wavelength of propagating signals ($a k_i \ll 2\pi, \  i\in\{o,e\}$), and far below the JJ plasma frequency, the dynamics of the node-flux in the even and odd modes can be described by a set of coupled PDEs \cite{supp},
\begin{eqnarray}\label{ode0}
\Phi_{tt}^{(i)} - \omega_{0i}^2 \Phi_{xx}^{(i)}
=\beta_{ij}^2 ( m \Phi_{xx}^{(j)} + m_x \Phi_x^{(j)}),
\end{eqnarray}
where we keep only the first order terms in $m=m(x,t)$ and assume $a=1$ for brevity. Subscript $t$ and $x$ represent partial derivatives of time and position respectively. Here $\beta_{ij}^2=\omega_{0i}^2 L_{dc}/L_j$, where $\{i,j\}\in\{ e,o \} , \ i \neq j$.
We recast Eq.\,\ref{ode0} in terms of forward/backward traveling waves in even and odd modes  $E_{f/b}(x,t) = (V_e \pm Z_e I_e)/2$ , $O_{f/b}(x,t) = (V_o \pm Z_o I_o)/2$ where $V_i= \Phi_t^{(i)}$ and $I_i= - \frac{1}{L_i} \Phi^{(i)}_{x} - \frac{1}{L_i L_j} \mathbb{L}_m(x,t) \Phi^{(j)}_x$ and consider ansatze in form of $E_{f/b}(x,t)= \mathcal{N}_e^{1/2} E(x) e^{-i(\omega_e t \mp k_e x)}+c.c$ , $O_{f/b}(x,t)=\mathcal{N}_o^{1/2} O(x) e^{-i(\omega_o t \mp k_o x)}+ c.c$. Here we conveniently choose scaling parameters $\mathcal{N}_{i}=Z_i \omega_i \, , i\in\{e,o\}$ so that $|E_{f/b}(x)|^2$ and $|O_{f/b}(x)|^2$ are proportional to the average photon number propagating in each mode. 
For the induced coupling parameter, we consider $m(x,t)=m(x) e^{-i(\omega_p t + \int k_p(x) dx)} + c.c$ where $\omega_p = \omega_o - \omega_e$.  Substituting these relations into the coupled equations (Eq.\,\ref{ode0}), and by applying the rotating wave approximations both in time and space \footnote{Basically we ignore fast oscillatory terms either due to frequency mismatch or wavevector mismatch. With these approximations, forward and backward dynamics become effectively decoupled. In supplemental material \cite{supp} (Fig.\,S3) we show that even with including higher oscillatory terms in the dynamical matrix we get similar result.} and keeping only the first order terms in $m(x)$, we arrive at a coupled wave equations $\mathbb{U}_x=\mathbb{M}.\mathbb{U}$ where $\mathbb{U}= [ E_f(x), O_f(x), E_b(x), O_b(x)]^{T}$ and the dynamical matrix (see supplemental material for detailed derivations) 
\begin{eqnarray}\label{ode4by4}
\mathbb{M}= \frac{i}{2}\begin{pmatrix}
- \Delta k_f(x) &   \kappa(x) & 0 & 0  \\
 \kappa(x) &  \Delta k_f(x)  & 0 & 0 \\
 0 & 0 &  \Delta k_b(x) &   \kappa(x)  \\
 0 & 0 & \kappa(x) & - \Delta k_b(x)  
\end{pmatrix},
\end{eqnarray}
where the phase mismatches are defined as in Eq.\,\ref{ode00}. Here the effective coupling is
\begin{equation}\label{kapxkeko}
\kappa(x) = \sqrt{\omega_e \omega_o /(Z_e Z_o)} L_{dc} m(x)\simeq \sqrt{k_e k_o} m(x),
\end{equation}
which is frequency dependent  unlike our simple presented model (Eq.\,\ref{ode00}).

 As illustrated in Fig.\,\ref{fig2}a, the spatially linear variation in pump wavevector is implemented by changing the capacitance and the inductance together in the pump transmission line as $k_p \propto\sqrt{L_p C_p}$. In this way, the impedance of the line remains constant as $Z_p\simeq\sqrt{L_p/C_p}$. With the linear variation of the pump wavevector, the phase mismatch would be a linear function of position $x$ (See Eq.\,\ref{LCS}). We choose circuit parameters so that with pump frequency $\omega_p/2\pi=$ 2 GHz,  perfect phase mismatching condition occurs at the center of the device for 6 GHz input signal which means in Eq.\,\ref{LCS} we set $K=k_o(6~ \mathrm{GHz} + 2~\mathrm{GHz}) - k_e(6~\mathrm{GHz})$, and for the pump wavevector sweep range we set $\alpha=0.05\ ( a^{-1})$ which is reasonably achievable in practice.


\begin{figure}[]
\centering
\includegraphics[width=0.98 \linewidth]{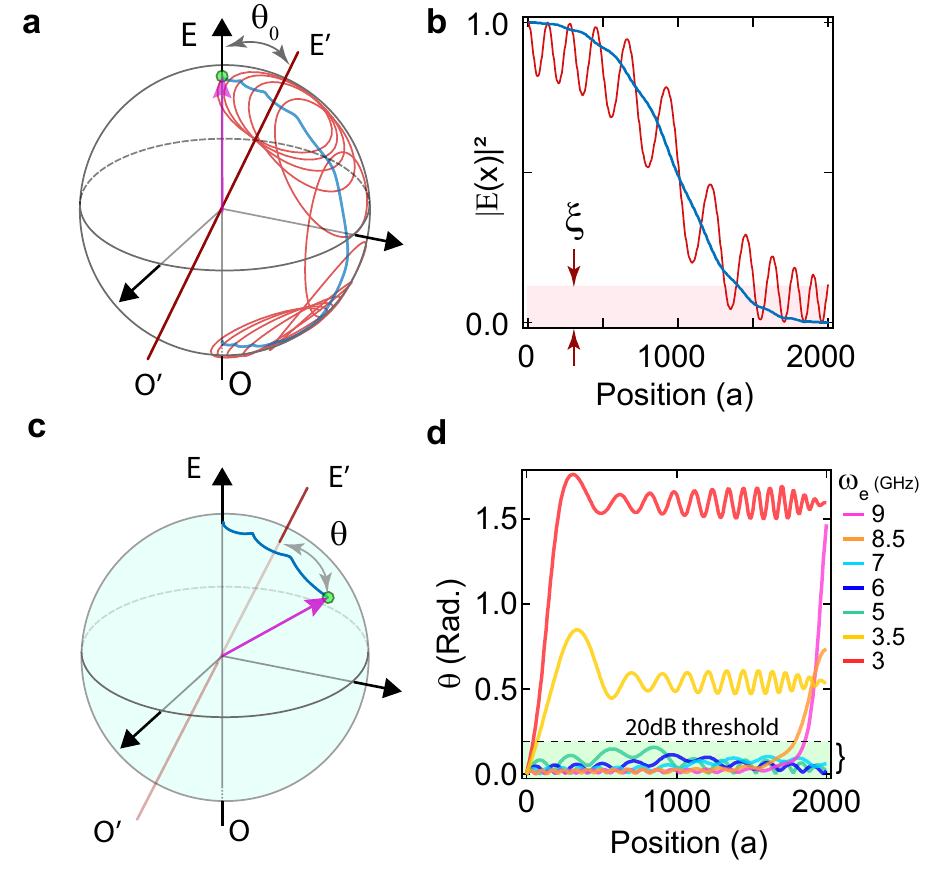}
\caption{\textbf{Bloch sphere representation of the adiabatic conversion:} \textbf{a.} E and O represent the even and odd mode which are the natural eigenmodes in the absence of the rf pump. E' and O' are eigenmodes in presence of rf pump. $\theta_0= \tan^{-1}[ \kappa(x=0)/\Delta k_f(x=0)]$ is the angle between the even mode and system's temporal eigenmode. The red (blue) curve is the adiabatic dynamics without (with) spatial coupling variation. \textbf{b.} The corresponding conversion performance without (with) spatial coupling variation shown in red (blue). The residual signal power in the even mode at $x=L$ can be estimated by $\xi =  \kappa_0^2/( \kappa_0^2 +  \alpha^2)\simeq 0.19$ which sets a lower-bound of -7 dB for the isolation. \textbf{c.} The angle $\theta$ is the angle between the actual state and the instantaneous eigenmode of the system during the conversion which quantifies the deviation from the full adiabatic evolution. \textbf{d.}  The deviation angle $\theta$ vs position for different signal frequencies. The deviation at the end of the device should be less than 0.2 [rad] (dashed line) to have 20\,dB isolation. \label{fig3}}
\end{figure}

Now we discuss the implementation of spatial coupling strength ramp (Eq.\,\ref{quadratic_ramp}) which is essential for this device to perform efficiently given practical limitations. The idea is that, to have an efficient adiabatic evolution, ideally, the system should start in an eigenmode and remain in its instantaneous eigenmode throughout the evolution. In our case, this requires that at $x=0$ the eigenmode of the system to be even/odd mode. Without spatial variation of coupling $\kappa(x)=\kappa_0$, the eigenmode of our system in the presence of flux drive would asymptotically approach to the even/odd mode in the limit of $\alpha\gg \kappa_0$. However, due to practical upper-bound limitation for the pump maximum wavevector sweep range $\alpha$ and lower-bound limitation for $\kappa_0$ due to the finite device length, this requirement would not be met (in our case $\alpha/\kappa_0\sim 2$). Therefore, as depicted in Fig.\,\ref{fig3}a and b, the signal starting at even mode at $x=0$ starts oscillating (the red path in \ref{fig3}a,b) and would not follow the ideal adiabatic path. The unwanted oscillation is translated to inefficient conversion at $x=L$, which limits the isolation performance.
However, increasing the mismatch sweep range $\alpha$ is not the only way to ensure that the system starts in the eigenmode.  One can slowly ramp off the coupling at both ends of the medium similar to pulse shaping techniques in the time domain \cite{martinis2014fast,guerin2011optimal,gambetta2011analytic,yang2017achieving}. With no coupling, mode E and O are eigenmodes and the signal starting at even mode at $x=0$ would not experience oscillatory evolution (e.g. the blue path in \ref{fig3}a,b). This technique of ramping the coupling drastically improves the adiabaticity of the evolution thus the efficiency of the conversion.
In our model discussed earlier, we consider a quadratic scaling for effective coupling to carry out ramp up/down (Eq.\,\ref{quadratic_ramp}). For the implementation we use slightly different ramping curve to partially compensate the frequency dependence of coupling (see Eq.\,\ref{kapxkeko}). In our adiabatic scheme, since the effective coupling for different frequency happens effectively at different position, we can partially compensate for the frequency dependency of the coupling by implementing slightly stronger coupling at lower $x$. We choose generalized Gaussian function,
\begin{equation}\label{quadratic_ramp_imp}
m(x)= \frac{m_0}{1-s} \left( e^{-\mid x/L-1/2\mid^p}- s \right)
\end{equation}
where the parameter $p=3$ for $x\leq L/2$ and $p=2$ for $x> L/2$ controls the ramp up (ramp down) steepness. The parameters $s=\exp(-1/2^{p})$ is a scaling factor to insure zero coupling at $x=0,L$ and maximum coupling at $x=L/2$. Here $m_0$ corresponds to the maximum modulated inductance. The implemented ramp function is depicted in Fig.\,\ref{fig1}c inset (green curve). Practically, the ramping implementation can be done by varying the effective mutual coupling between pump line and SQUID loops as schematically illustrated in Fig.\,\ref{fig2}a.

Therefore, we have all essential components for the implementation of the proposed adiabatic scheme with a dynamics governed by Eq.\,\eqref{ode4by4}. By terminating odd mode at both ends of the device (shown in Fig.\,\ref{fig1}a), and given the circuit parameters listed in Table\,1, it gives more than 20 dB isolation over an octave of bandwidth as depicted in Fig.\,\ref{fig1}d (green curve).

\emph{adiabaticity analysis}---We quantify the adiabaticity of the evolution by calculating the parameter $\theta$ which is the angle between the state of system (the magenta arrow in Fig.\,\ref{fig3}c) and the instantaneous eigenstate (red line in Fig.\,\ref{fig3}c) at any given position $x$. Ideally in a full adiabatic evolution, $\theta$ is zero throughout the conversion which means system always follows its instantaneous eigenstate. In Fig.\,\ref{fig3}d we plot $\theta$ for different signal frequencies which stays close to zero for signals near 6 GHz and deviates significantly for 4 and 8 GHz as we expected from result in Fig.\,\ref{fig1}d. The angle at the end of the device $\theta_L=\theta|_{x=L}$  determines the isolation as the probability of finding the signal at mode E at the end of the device is $P_e\simeq \theta_L^2/4$. The dashed line in Fig.\,\ref{fig3}d indicates $\theta_L=0.2\, (rad.)$ the threshold angle below which we get more than 20 dB isolation. 
It is worth mentioning that, the final angle $\theta_L$ can be estimated by a simple geometrical analysis which is used for error estimation in adiabatic two qubit gates \cite{martinis2014fast}, 
\begin{eqnarray}\label{geo_anly}
\theta_{L} = -\int_0^L (\frac{ d \theta_{adi} }{d x}) dx \exp[-i \smallint g(x') dx'],
\end{eqnarray}
where $\theta_{adi} = \arctan[ \kappa(x)/ \Delta k_f(x)]$ and $g(x)=\sqrt{\kappa(x)^2 + \Delta k_f(x) ^2}$. For 20 dB isolation, the probability of the excitation at the end of the adiabatic evolution should be at most 1\% ($P_e<0.01$) so $\theta_L$ should be less that 0.2 radian.

\begin{figure}[]
\centering
\includegraphics[width=0.98 \linewidth]{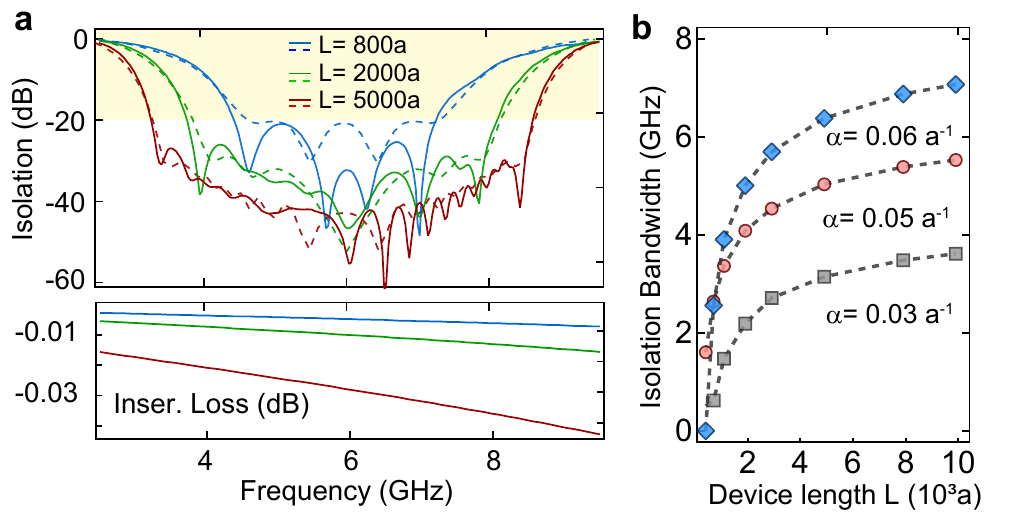}
\caption{\textbf{Isolation performance vs device length:} \textbf{a.} The green (red) dotted, dashed and solid line show the isolation (transmission) performance of the device with L=800a, 2000a and 5000a respectively. \textbf{b.} The square, circular and diamond markers with dashed lines as guides to the eye, show the isolation bandwidth versus device lengths for mismatch sweep range $\alpha=$ 0.03 $a^{-1}$, 0.05 $a^{-1}$, 0.06 $a^{-1}$ respectively.
\label{fig4}.}
\end{figure}

In Fig.\,\ref{fig4}a,we plot the isolation and transmission (insertion loss) performance for different device length $L$ (number of unit cells) given $\alpha=0.05\,a^{-1}$. Notably, the device with 800 unit cells provides over 2 GHz isolation bandwidth. The dashed lines are the estimated isolation using geometrical analysis (Eq.\,\ref{geo_anly}) which shows a great agreement with our result.
The corresponding insertion loss plotted in the lower sub-panel in Fig.\,\ref{fig4}a is obtained by calculating the conversion of the mode E to mode O in backward direction and considering the typical dielectric loss in the superconducting qubit fabrications (loss tangent $ \delta=10^{-5}$). 
Figure~\ref{fig4}c shows the isolation bandwidth versus the device length with given parameters in table~1 for three different mismatch sweep range $\alpha$. For infinitely long device the isolation approaches to the corresponding frequency range for pump wavevector sweep (e.g. for $\alpha=0.05$ the isolation bandwidth approaches to $6.6$ GHz which is the highlighted frequency range in Fig.\,\ref{fig1}b).

\emph{mode splitter and impedance considerations}--- The impedance of the even and odd modes can be independently set by circuit parameters to the desired value. In this letter, we choose the parameters so that each transmission lines have 50 $\Omega$ impedance.
However, the impedance of the device is ultimately mandated by the termination configuration and the choice for the even/odd mode splitter at both ends of the device. For example, the odd mode can be blocked by a built-in broadband Wilkinson power divider. Alternatively one can use a broadband 180 hybrid couplers that split even and odd modes.
The lumped element version of both types of terminations are readily available which can be naturally integrated with the device \cite{shen2009design,gao2013design}.

\emph{conclusion}--- Superconducting circuits have been shown a promising capability for realizing novel quantum devices. However, the realization of a large scale quantum system requires integration of its essential components. Here we show that nonlinear processes in coupled transmission lines open new possibilities of controlling photon dispersion and engineering novel quantum devices. We proposed a practical scheme for on-chip broadband isolator using adiabatic techniques and dispersion engineering in coupled superconducting transmission lines. The presented idea can be readily applied to other platforms such as silicon photonics.  

\emph{acknowledgment}---
This work was funded in part by the AWS Center for Quantum Computing and by the MIT Research Support Committee from the NEC Corporation Fund for Research in Computers and Communications. Y. Ye acknowledges financial support from the MIT EECS Jin Au Kong fellowship. G. Cunningham acknowledges support from the Harvard Graduate School of Arts and Sciences Prize Fellowship.
\bibliographystyle{unsrt}
\bibliography{main}

\end{document}


\title{Supplemental Material:\\ Broadband Microwave Isolation with Adiabatic Mode Conversion in Coupled Superconducting Transmission Lines}

\author{Mahdi Naghiloo}
\affiliation{Research Laboratory of Electronics, MIT,  Cambridge, Massachusetts 02139}

\author{Kaidong Peng}
\affiliation{Research Laboratory of Electronics, MIT,  Cambridge, Massachusetts 02139}
\affiliation{Department of Electrical Engineering and Computer Science, MIT,  Cambridge, Massachusetts 02139}

\author{Yufeng Ye}
\affiliation{Research Laboratory of Electronics, MIT,  Cambridge, Massachusetts 02139}
\affiliation{Department of Electrical Engineering and Computer Science, MIT,  Cambridge, Massachusetts 02139}

\author{Gregory Cunningham}
\affiliation{Research Laboratory of Electronics, MIT,  Cambridge, Massachusetts 02139}
\affiliation{Harvard John A. Paulson School of Engineering and Applied Sciences, Harvard University, Cambridge, MA 02138, USA}

\author{Kevin P. O'Brien}
\affiliation{Research Laboratory of Electronics, MIT,  Cambridge, Massachusetts 02139}
\affiliation{Department of Electrical Engineering and Computer Science, MIT,  Cambridge, Massachusetts 02139}




\maketitle


\section{Circuit diagram}
\renewcommand{\thesection}{S\arabic{section}}  
\renewcommand{\thetable}{S\arabic{table}}  
\renewcommand{\thefigure}{S\arabic{figure}}
\renewcommand{\theequation}{S\arabic{equation}}
\setcounter{page}{1}
\setcounter{figure}{0}
\setcounter{equation}{0}

\begin{figure}[h]
\centering
\includegraphics[width=5 in]{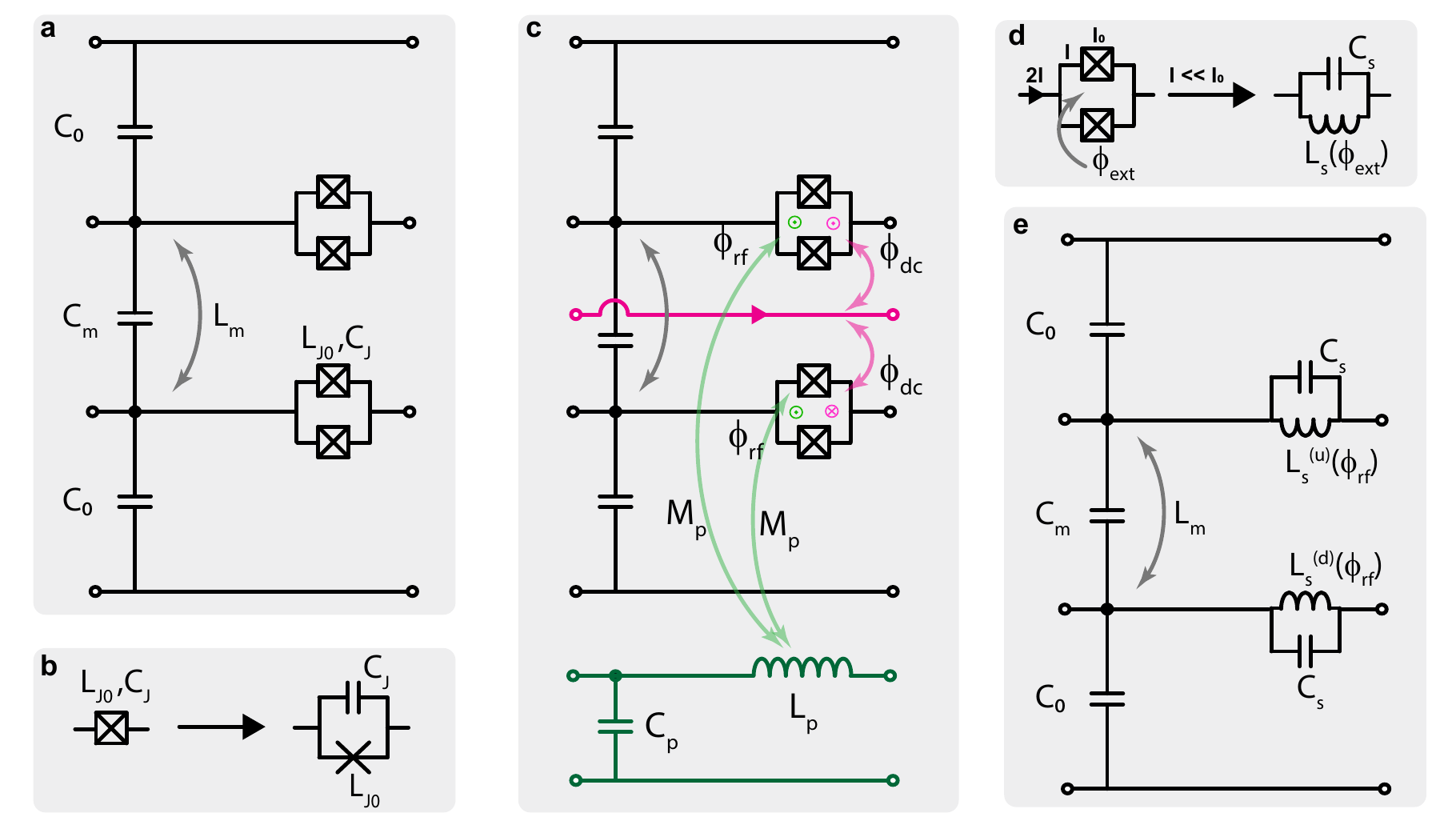}
\caption{\textbf{Circuit diagrams}  \label{fig1s}}
\end{figure}

Figure\,\ref{fig1s}, summarizes the details about the circuit diagram of the coupled Josephson junction transmission line that have been considered in this work. Panel a (Fig.\,\ref{fig1s}\,a)  shows typical unit cell of two identical coupled transmission lines. The inductance in each line is predominantly provided by a pair of Josephson Junctions (JJ) forming a SQUID. Each JJ is characterized by its inductance $L_{J0}$ and the shunted capacitance $C_j$ as depicted in the panel (b). The JJ inductance is directly related to the critical current of the junction  $L_{J0}= \phi_0/I_0$ where $\phi_0= \hbar/(2e)$ is reduced flux quantum. 
Here $C_0$ is the capacitance to the ground and $C_m$ and $L_m$ are mutual capacitance and inductance between two transmission lines.

In each unit-cell, a pair of identical JJs form a SQUID whose effective inductance $L_s$ depends on current $I$ passing each JJ and the external fluxes passing the SQUID loop,
\begin{eqnarray}\label{squid_L_I_phi}
L_s(I, \phi_{ext}) = \frac{L_{J0}}{2} \frac{1}{\sqrt{1-I^2/I_0^2}} \frac{1}{ | \cos(\frac{\phi_{ext}}{2 \phi_0}) | }.
\end{eqnarray}

As depicted in panel c (Fig.\,\ref{fig1s}\,c) Each SQUID receive a DC flux bias (shown in magenta) as well as a rf flux from a pump transmission line shown in green. The DC and rf pump transmission lines are linearly decoupled from the coupled transmission lines since they have different dispersion relations. In such situation, the entire effect of DC and rf transmission lines can be captured by their induced fluxes on the SQUID loops. We consider a configuration that in each unit cell the upper and lower SQUID receive DC bias in opposite direction but the rf flux is in-phase for both SQUIDs as illustrated in panel c (Fig.\,\ref{fig1s}\,c). Note that although in our illustration the pump transmission line has different distance to SQUID loops, in actuality the situation is symmetric and the effective mutual inductance between pump and SQUID loops $M_p$ are the same for both lines. Therefore the magnitude of the induces flux $\phi_{rf}$ is equal for both upper and lower SQUIDs within a unit cell.

As shown in Eq.\, \ref{squid_L_I_phi}, the inductance of SQUID also depends on the current passing through the JJs. However, we consider weak signal regime where the current associated with propagating signals in upper/lower transmission lines are much smaller that critical current of JJs ($I \ll I_0 $) so that the nonlinear effects due to propagating currents are negligible. In such situation, the SQUID can be considered as an inductor whose inductance is modulated by external flux as shown in panel\,d (Fig.\,\ref{fig1s}\,d). Therefor, the effective inductance in upper and lower transmission line would be,

\begin{eqnarray}\label{ind_L_m}
L_s^{(u),(d)} = \frac{ 0.5 L_{J0}  }{| \cos(\frac{\phi_{dc}}{2 \phi_0} \mp \frac{\phi_{rf}}{2 \phi_0}) | } \simeq \frac{ 0.5 L_J  }{ \cos(\frac{\phi_{dc}}{2 \phi_0}) \pm \sin(\frac{\phi_{dc}}{2 \phi_0} ) \frac{\phi_{rf}}{2 \phi_0}} = \frac{L_{dc}}{1 \pm m}.
\end{eqnarray}
Where the unitless parameter $m=\tan(\frac{\Phi_{dc}}{2 \phi_0}) \frac{\phi_{ac}}{2 \phi_0}$ accounts for flux pumped induced modulation and $L_{dc}=0.5 L_{J0}/\cos(\frac{\phi_{dc}}{2\phi_0})$ is the DC-biased SQUID inductance. We consider DC flux bias in each SQUID to be  $\phi_{dc}=2 \pi/3 \ \phi_0$ so we have $L_{dc}=L_{J0}$.

The final circuit model of a typical unit cell is shown in panel e (Fig.\, \ref{fig1s}\,e). The proposed device can be modeled by cascading many of these unit cells. Note that all the circuit parameters are fixed throughout the device except the pump transmission line characteristics. We let $L_p$ and $C_p$ and the inductive coupling $Mp$ slowly vary versus position (unit cell number). These spatial variations are slow in the sense that the variations between two adjacent unitcell is negligible. These variations of pump characteristics are all captured in $\phi_{rf}$ in our model as shown in panel e (Fig.\, \ref{fig1s}\,e) .

\section{Lagrangian analysis}

\begin{figure}[h]
\centering
\includegraphics[width=3.5 in]{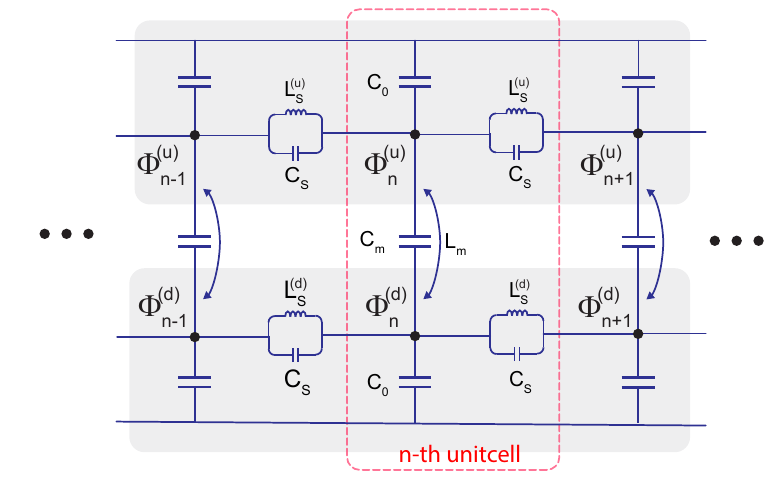}
\caption{\textbf{Node fluxes}  \label{node_flux}}
\end{figure}

The Lagrangian for a coupled transmission line can be written in terms of node-fluxes $\Phi_n^{(u)}$ and $\Phi_n^{(d)}$ (shown in Fig\,\ref{node_flux}).

For n-th unit cell, the kinetic and potential energy would be
\begin{eqnarray}
T_n&=&
+ \frac{1}{2} C_0 \left( \dot{\Phi}_n^{(\mathrm{u})}\right)^2
 + \frac{1}{2} C_s   \left( \dot{\Phi}_{n+1}^{(\mathrm{u})} -\dot{\Phi}_n^{(\mathrm{u})} \right)^2
 +\frac{1}{2} C_0 \left( \dot{\Phi}_n^{(\mathrm{d})}\right)^2
 + \frac{1}{2} C_s   \left( \dot{\Phi}_{n+1}^{(\mathrm{d})} -\dot{\Phi}_n^{(\mathrm{d})} \right)^2
+ \frac{1}{2} C_m \left(\dot{\Phi}_{n}^{(\mathrm{u})} -\dot{\Phi}_n^{(\mathrm{d})} \right)^2,\\
 U_n&=&
 +\frac{1}{2} \frac{1}{L_{s}^{\mathrm{(u)}}L_{s}^{\mathrm{(d)}} - L_m^2} \left[ L_{s}^{\mathrm{(d)}}   (  \Phi_{n+1}^{(\mathrm{u})} - \Phi_{n}^{(\mathrm{u})})^2 + L_{s}^{\mathrm{(u)}}   (  \Phi_{n+1}^{(\mathrm{d})} - \Phi_{n}^{(\mathrm{d})} )^2 - 2L_m ( \Phi_{n+1}^{(\mathrm{u})} -\Phi_{n}^{(\mathrm{u})} )  ( \Phi_{n+1}^{(\mathrm{d})} -\Phi_n^{(\mathrm{d})} )  \right].
\end{eqnarray}
Where, for brevity, we dropped the explicit time dependencies for $\Phi_{n}^{(\mathrm{u,d})}$ and also $\phi_{ext}$ dependence for $L_s^{(\mathrm{u,d})}$. 
As we discussed in previous section, the inductance of SQUIDs in upper an lower transmission lines are modulated in opposite direction around $L_{dc}$ (See Eq. \ref{ind_L_m}). By considering Eq. \ref{ind_L_m} and using subscript $n$ for n-th unit cell we rewrite the potential energy as

\begin{eqnarray}
 U_n&=&
 \frac{1}{2} \frac{L_{dc}}{L_{dc}^2 - L_m^2} \left[  (1+m_n)   \left(  \Phi_{n+1}^{(\mathrm{u})} - \Phi_{n}^{(\mathrm{u})} \right)^2 +  (1-m_n)  \left(  \Phi_{n+1}^{(\mathrm{d})} - \Phi_{n}^{(\mathrm{d})} \right)^2 - \frac{2 L_m}{L_{dc}} \left( \Phi_{n+1}^{(\mathrm{u})} -\Phi_{n}^{(\mathrm{u})} \right)  \left( \Phi_{n+1}^{(\mathrm{d})} -\Phi_n^{(\mathrm{d})} \right)  \right]. \nonumber 
\end{eqnarray}
Where we keep terms in the first order of $m_n$ (weak modulation regime). We recast the kinetic and potential energies in the even and odd modes basis considering the following definition
\begin{eqnarray}
\Phi^{(i)}_n = \frac{\Phi^{(u)}_n \pm \Phi^{(d)}_n}{\sqrt{2}} \ , \ i \in \{ e, o\}
\end{eqnarray}
Then we have,
\begin{eqnarray}
T_n&=&
+ \frac{1}{2} C_e \left( \dot{\Phi}_n^{(e)}\right)^2
 %
 %
 +\frac{1}{2} C_0 \left( \dot{\Phi}_n^{(o)}\right)^2 
 ,\\
U_n&=&
\frac{1}{2 L_{e}} \left( \Phi_{n+1}^{(e)}  - \Phi_{n}^{(e)} \right)^2 +  \frac{1}{2 L_{o}} \left( \Phi_{n+1}^{(o)}  - \Phi_{n}^{(o)} \right)^2 + \frac{\mathbb{L}_m}{L_{e} L_{o}} \left( \Phi_{n+1}^{(e)}  - \Phi_{n}^{(e)} \right) \left( \Phi_{n+1}^{(o)}  - \Phi_{n}^{(o)} \right)  .
\end{eqnarray}
where we keep only up to the first order in $\mathbb{L}_m = L_{dc} m_n$. Here $C_e=C_0$, $C_o= C_0+2 C_m$, $L_e=L_{dc}+L_m$, and $L_o= L_{dc}-L_m$ are corresponding capacitors and inductance in even and odd modes as depicted in the main text (Fig.\,2\,d).
Note that we ignore the effect of $C_s$ as we have $\omega^2 L_{dc} C_s \ll 1 $ in the frequency range of interest.
The full Lagrangian for the system would be  $ \mathcal{L} = \sum_{n=1}^{N} \mathcal{L}_n $ where $ \mathcal{L}_n = T_n - U_n $. Now we use Euler-Lagrange relations 
\begin{eqnarray}
\frac{\ \partial \,\mathcal{L}\ \ \ }{\partial \Phi_{n}^{(\mathrm{i})} }=\frac{d}{dt} \left( \frac{\ \partial \,\mathcal{L}\ \ \ }{\partial \dot{\Phi}_{n}^{(\mathrm{i})}} \right).
\end{eqnarray}
to arrive equations-of-motion for node fluxes,
\begin{small}
\begin{eqnarray}
\frac{1}{L_{i}}   \left( 2 \Phi_{n}^{(i)} - \Phi_{n+1}^{(i)} - \Phi_{n-1}^{(i)} \right)
 +\frac{L_{dc}}{L_{i} L_j} \left(   - m_n ( \Phi_{n+1}^{(i)} - \Phi_{n}^{(i)}) + m_{n-1} ( \Phi_{n}^{(i)} - \Phi_{n-1}^{(i)})  \right)=  C_i  \ddot{\Phi}_n^{(i)}.
\end{eqnarray}
\end{small}

Note that the conjugate momentum (node-charge) associated with the node-flux can be calculates as
\begin{eqnarray}\label{nodecharges}
\Pi^{(i)}_n = \frac{\partial }{\partial \dot{\Phi}^{(i)}_n} \mathcal{L} = C_i \dot{\Phi}^{(i)}_n
\end{eqnarray}
To proceed further, we go to the continuum limit (assuming $a k_i \ll 2 \pi,\  i=\{e,o,p\}$) and drop subscript $n$ and define derivatives (e.g.  $  \Phi_n - \Phi_{n-1}\rightarrow a  \Phi_{x}$ and $   \Phi_{n+1}   + \Phi_{n-1} - 2 \Phi_n  \rightarrow a^2  \Phi_{xx}$  ) where $a$ is the unit cell length. Hereafter, we use the convention that subscripts $x$ and $t$ represents the derivatives.
In the continuum limit we have
\begin{eqnarray}\label{ode0s_lag}
\Phi^{(i)}_{tt} - \omega_{0i}^2 \Phi^{(i)}_{xx} 
=\beta_{ij}^2 ( m_x \Phi^{(j)}_x + m \Phi^{(j)}_{xx} ).
\end{eqnarray}
Here $\omega_{0i}=1/\sqrt{L_i C_i}$\, and  $\beta^2_{ij} = \omega_{0i} L_{dc}/ L_j$ where $i \in \{ e ,o\} , i \neq j$ and we set $a=1$ for brevity.
We characterize the linear property of the system by analyzing Eq.\,\ref{ode0s_lag} in the absence of the pump modulation ($m \rightarrow 0$). In this situation the right hand sides of \ref{ode0s_lag} vanishes and the remaining equations describes linear property of the system which can be solved to obtain dispersion relations of each mode. For that, we assume $\Phi^{(i)}=\Phi^{(i)}(x,t)=\phi^{(i)}(0) e^{-i(\omega_i t - k_i x))} + c.c$ and obtain,
\begin{eqnarray}\label{KEKO_s}
k_i= \frac{\omega_i/\omega_{0i}}{\sqrt{1- (\frac{\omega_i}{\omega_{0J}})^2 }}\simeq \omega_i/\omega_{0i} = \sqrt{L_i C_i} \, \omega_i \  , \  i \in \{ e , o ,\}.
\end{eqnarray}
Here $\omega_{0i}=1/\sqrt{C_i L_i}$, and $\omega_{Ji}=1/\sqrt{C_s L_i}$ are cut-off frequencies and plasma frequency respectively.

In order to proceed with nonlinear analysis, it would be convenient to rearrange Eq.\ref{ode0s_lag} as,
\begin{eqnarray}\label{ode0s_lag_r}
\partial_t \left( \Phi^{(i)}_{t} \right)= \partial_x \left( \omega_{0i}^2 \Phi^{(i)}_{x} + \beta_{ij}^2  m(x,t) \Phi^{(j)}_x \right).
\end{eqnarray}
%
%
%
Now we define the voltage $V_i = \Phi^{(i)}_t $ and current $I_i= - \frac{1}{L_i} \Phi^{(i)}_{x} -  \frac{L_{dc}}{L_i L_j}  m(x,t) \Phi^{(j)}_x$ and rearrange terms (keeping terms up to the first order in $m(x,t)$) to have,

\begin{eqnarray}
\frac{\partial}{\partial x} V_i &=& - L_i \dot{I}_i +  \frac{\partial }{\partial t} \left( \mathbb{L}_m(x,t) I_j \right),\\
\frac{\partial }{\partial x} I_i &=&  -  C_i \dot{V}_i   ,
\end{eqnarray}
%
%
where, $\mathbb{L}_m(x,t)= L_{dc} m(x,t)$ is the induced-inductive coupling. Therefore, we arrived at equations of motion consistent with the coupled mode theory analysis \cite{orfanidis2002electromagnetic}. For clarity we write down all four ODEs explicitly,

\begin{subequations}\label{odesVI}
\begin{eqnarray}
\frac{d}{dx} V_e(x,t) &=& -  L_{e} \frac{d}{dt} I_e(x,t)  +   \frac{d}{dt} \left[ \mathbb{L}_m(x,t) I_o(x,t) \right] \\
\frac{d}{dx} V_o(x,t) &=& -  L_{o} \frac{d}{dt} I_o(x,t)  +   \frac{d}{dt} \left[ \mathbb{L}_m(x,t) I_e(x,t) \right] \\
\frac{d}{dx} I_e(x,t) &=& -  C_e \frac{d}{dt} V_e(x,t) \\
\frac{d}{dx} I_o(x,t) &=& -  C_o \frac{d}{dt} V_o(x,t),
\end{eqnarray}
\end{subequations}
Therefore, in even and odd mode basis we have two  otherwise decoupled transmission lines where the inductive coupling $\mathbb{L}_m(x,t)=L_{dc}m(x,t)$ is induced by external flux pump.

Now we recast the voltage and current in form of forward and backward propagating signals given the following relationships,
\begin{subequations} \label{FB_waves}
\begin{eqnarray}
E_{f/b}(x,t) = \frac{V_e(x,t) \pm Z_e I_e(x,t)}{2} ,  \\
O_{f/b}(x,t) = \frac{V_o(x,t) \pm Z_o I_o(x,t)}{2},
\end{eqnarray}
\end{subequations}
where $Z_i=\sqrt{L_i/C_i}, i \in \{e,o \}$. In order to proceed, we consider the flowing ansatze for even and mode signal propagating in both directions
\begin{subequations}\label{EO_anzats}
\begin{eqnarray}
E_{f/b}(x,t) = E_{f/b}(x) e^{-i(\omega_e t \mp k_e x)}+ c.c.\\
O_{f/b}(x,t) = O_{f/b}(x) e^{-i(\omega_o t \mp k_o x)}+c.c.,
\end{eqnarray}
\end{subequations}
 and consider a traveling wave form for the induced coupling
\begin{equation}\label{mxt}
\mathbb{L}_m(x,t)= L_{dc}m(x,t)=L_{dc}m(x)e^{-i( \omega_p t - \int k_p dx)} + c.c.
\end{equation}
where $k_p$ is the instantaneous pump wave vector and $\int k_p dx$ accounts for accumulated phase due to pump phase wavevector. Substituting Eq\,\ref{EO_anzats} and Eq.\,\ref{mxt} in Eq.\,\ref{odesVI}, we have the following form for the dynamical matrix $\mathbb{M}$,
\begin{small}
\begin{eqnarray} 
\mathbb{M}
=
\begin{pmatrix}
0 &  \frac{i L_{dc} m(x) \omega_e e^{-i \Delta K_f(x)}}{2Z_o} & 0 & -\frac{iL_{dc} m(x) \omega_e e^{-i (\Delta K_f(x) - 2 k_o x )}}{2Z_o} \\
 \frac{i L_{dc} m^*(x) \omega_o e^{i \Delta K_f(x)}}{2Z_e}&  0 &-\frac{i L_{dc} m^*(x) \omega_o e^{i (\Delta K_f(x)+2k_e)}}{2Z_e}& 0 \\
0 & \frac{i L_{dc} m(x) \omega_e e^{-i (\Delta K_f(x) + 2 k_e x )}}{2Z_o}& 0 & -\frac{i L_{dc} m(x) \omega_e e^{-i (\Delta K_f(x) - 2 (k_o-k_e) x )}}{2Z_o} \\
\frac{i L_{dc} m^*(x) \omega_o e^{i (\Delta K_f(x)-2k_o)}}{2Z_e} & 0 & -\frac{i L_{dc} m^*(x) \omega_o e^{i (\Delta K_f(x)-2(k_o-k_e)x)}}{2Z_e}& 0
\end{pmatrix}
\end{eqnarray}
\end{small}
where we applied RWA in time domain. Here $\Delta K_f(x) = (k_o-k_e)x -\int{k_p dx}$ is accumulated phase mismatch. We may recast the obtained coupled equations it in a rotating frame corresponding to the following transformation $E_{f}(x) \rightarrow E_{f}(x) e^{-i \Delta K_f(x) /2}, O_{f}(x) \rightarrow  O_{f}(x) e^{+i  \Delta K_f(x) /2}$ and $E_{b}(x) \rightarrow E_{b}(x) e^{-i \Delta K_f(x)/2+(k_o-k_e)x}, O_{b}(x) \rightarrow  O_{b}(x) e^{+i  \Delta K_f(x) /2 - (k_o-k_e)x}$ to obtain,

\begin{small}
\begin{eqnarray} \label{full4by4}
\mathbb{M}
=
\begin{pmatrix}
-\frac{i}{2}\Delta k_f(x) &  -\frac{i L_{dc} m(x) \omega_e }{2Z_o} & 0 & \frac{iL_{dc} m(x) \omega_e }{2Z_o} e^{i  (k_o+k_e) x } \\
 -\frac{i L_{dc} m(x) \omega_o }{2Z_e}&  \frac{i}{2} \Delta k_f(x) &\frac{i L_{dc} m(x) \omega_o e^{i (k_o+k_e)}}{2Z_e}& 0 \\
0 & -\frac{i L_{dc} m(x) \omega_e e^{-i (k_o+k_e) x }}{2Z_o}& -\frac{i}{2} \Delta k_f(x)+i(k_o-k_e) & \frac{i L_{dc} m(x) \omega_e }{2Z_o} \\
-\frac{i L_{dc} m(x) \omega_o e^{-i (k_o+k_e)}}{2Z_e} & 0 & \frac{i L_{dc} m(x) \omega_o}{2Z_e}& \frac{i}{2}\Delta k_f(x) - i(k_o-k_e)
\end{pmatrix}.
\end{eqnarray}
\end{small}

We may also simplified the equations even further to show its correspondence to the simple model presented in the main text. For that, we perform RWA in space and ignore highly oscillatory terms in space. With this approximation the forward and backward dynamics decouple.

\begin{small}
\begin{eqnarray} \label{simple4by4}
\mathbb{M}
=
\begin{pmatrix}
-\frac{i}{2}\Delta k_f(x) &  -\frac{i L_{dc} m(x) \omega_e }{2Z_o} & 0 & 0 \\
 -\frac{i L_{dc} m(x) \omega_o }{2Z_e}&  \frac{1}{2} \Delta k_f(x) &0& 0 \\
0 &  0 & -\frac{i}{2} \Delta k_f(x)+i(k_o-k_e) & +\frac{i L_{dc} m(x) \omega_e }{2Z_o} \\
0 & 0 & \frac{i L_{dc} m(x) \omega_o}{2Z_e}& \frac{i}{2}\Delta k_f(x) - i(k_o-k_e)
\end{pmatrix}
\end{eqnarray}
\end{small}

It is convenient to represent signals in terms of photon number basis which is conserved during the conversion for signals (tracing out the pump). So we apply the following transformation to go from voltage signal to photon number,

$E_{f/b}(x) \rightarrow 1/\sqrt{Z_e \omega_e} E(x), O_{f/b}(x) \rightarrow 1/\sqrt{Z_o \omega_o} O_{f/b}(x) $ where now $| E_{f/b}(x)|^2$ and $| O_{f/b}(x)|^2$ are proportional to average number of photon propagating in each mode.

\begin{small}
\begin{eqnarray} 
\frac{d}{dx}\begin{pmatrix}
E_f(x) \\
O_f(x)\\
E_b(x)\\
O_b(x)
\end{pmatrix}
=
\begin{pmatrix}
-\frac{i}{2}\Delta k_f(x) &  -\frac{i L_{dc} m(x) \sqrt{\omega_e \omega_o} }{2\sqrt{Z_e Z_o}} & 0 & 0 \\
 -\frac{i L_{dc} m(x) \sqrt{\omega_e \omega_o} }{2\sqrt{Z_e Z_o}}&  \frac{1}{2} \Delta k_f(x) &0& 0 \\
0 &  0 & -\frac{i}{2} \Delta k_f(x)+i(k_o-k_e) & \frac{i L_{dc} m(x) \sqrt{\omega_e \omega_o}}{2\sqrt{Z_e Z_o}} \\
0 & 0 & \frac{i L_{dc} m(x) \sqrt{\omega_e \omega_o}}{2\sqrt{Z_e Z_o}}& \frac{i}{2}\Delta k_f(x) - i(k_o-k_e)
\end{pmatrix}
.
\begin{pmatrix}
E_f(x) \\
O_f(x)\\
E_b(x)\\
O_b(x)
\end{pmatrix}
\end{eqnarray}
\end{small}

The obtained equation is corresponds to the simple model introduced in the main text with $\kappa(x) = \frac{\sqrt{\omega_e \omega_o}}{\sqrt{Z_e Z_o}} L_{dc} m(x)$ and $\Delta k_b(x) =   2(k_o -k_e)-\Delta k_f(x)$ which results in a poor phase mismatch for backward traveling signals.

\section{Spatial rotating wave approximation verification}
In the main text we perform spatial RWA and ignore the fast spatial oscillatory terms in the dynamical matrix (Eq.\,\ref{simple4by4}, or Eq.\,9 in the main text). With this approximation the forward and backward signals decouple as introduced in the simple model. Here we verify that this approximation is indeed valid by solving the full dynamical matrix  (Eq.\,\ref{full4by4}) and comparing the result with the solution of the Eq.\,\ref{simple4by4}.

\begin{figure}[h]
\centering
\includegraphics[width=3 in]{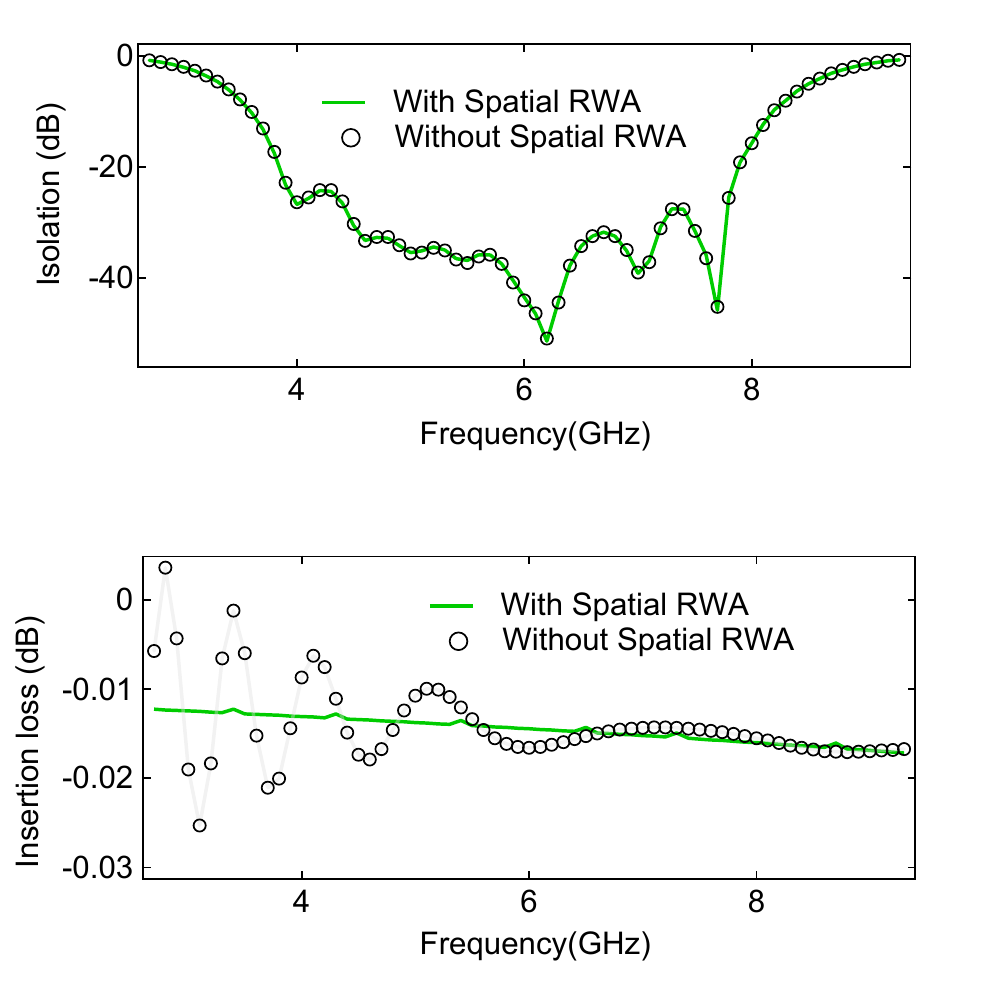}
\caption{\textbf{Isolation performance with/without spatial RWA} The upper (lower) panel compare the transmission in forward (backward) direction with/without performing spatial rotating wave approximation. \label{RWA_verigication}}
\end{figure}